\documentclass[preprint,showpacs,preprintnumbers,amsmath,amssymb,prb]{revtex4-1}

\usepackage{graphicx}

\begin{document}

\title{High-field magnetization and magnetoresistance 
of the $A$-site ordered perovskite oxide 
CaCu$_{3}$Ti$_{4-x}$Ru$_{x}$O$_{12}$~($0 \le x \le 4$)}

\author{T. Kida} \email{kida@mag.cqst.osaka-u.ac.jp}

\author{R. Kammuri}
 \altaffiliation[present affiliation:]%
{CASIO Computer Co.,Ltd., 1-6-2 Honmachi, Shibuya, Tokyo 151-8543, Japan}

\author{M. Hagiwara}
 \affiliation{KYOKUGEN, Osaka University, 
1-3 Machikaneyama, Toyonaka, Osaka 560-8531, Japan}

\author{S. Yoshii}
 \affiliation{Institute for Materials Research, 
Tohoku University, 2-1-1 Katahira, Sendai, Miyagi 980-8577, Japan}

\author{W. Kobayashi}
 \altaffiliation[present affiliation:]%
{Department of Physics, University of Tsukuba, 
1-1-1 Tennodai, Tsukuba 305-8571, Japan}
 \affiliation{WIAS, Waseda University, 
1-6-1 Nishiwaseda, Shinjuku, Tokyo 169-8050, Japan}

\author{M. Iwakawa}
 \affiliation{Department of Applied Physics, 
Waseda University, 3-4-1 Okubo, Shinjuku, Tokyo 169-8555, Japan}

\author{I. Terasaki}
 \affiliation{Department of Physics, Nagoya University, 
Furocho, Chikusa, Nagoya 464-8602, Japan}
\date{\today}

\begin{abstract}
We have measured high-field magnetization and magnetoresistance 
of polycrystalline samples of the $A$-site ordered perovskite 
CaCu$_{3}$Ti$_{4-x}$Ru$_{x}$O$_{12}$ ($0 \le x \le 4$) 
utilizing a non-destructive pulsed magnet. 
We find that the magnetization for $x=0.5$,  1.0 and 1.5 is
nonlinear, and tends to saturate in high fields. 
This is highly nontrivial because 
the magnetization for $x=0$ and 4 is linear in external field up 
to the highest one. 
We have analyzed this field dependence 
based on the thermodynamics of magnetic materials, 
and propose that the external fields 
delocalize the holes on the Cu$^{2+}$ ions 
in order to maximize the entropy. 
This scenario is qualitatively consistent with a large magnetoresistance 
of  $-$70~\% observed at 4.2~K at 52~T for $x=$1.5. 
\end{abstract}

\maketitle

%
\section{Introduction}
Filling-control metal-insulator transition 
on transition-metal oxides and their derivatives, 
such as high-$T_{\rm c}$ superconducting cuprates 
and colossal magnetoresisitive manganites, 
has attracted much attention in recent years 
because of their unique physical properties arising 
from interacting degrees of freedom between spin, orbital, 
charge, and lattice.\cite{dagotto2005} 
The layered perovskite oxide La$_{2-x}$Sr$_{x}$CuO$_{4}$ is 
a typical example, 
where Sr$^{2+}$ substitution for La$^{3+}$ 
works as an acceptor to supply a hole.\cite{takagi1989} 
The system evolves from an antiferromagnetic insulator 
to a paramagnetic metal with increasing $x$, 
where the high-temperature superconductivity emerges 
between the antiferromagnetic and paramagnetic phases. 
From a microscopic viewpoint, 
the electronic configuration of a Cu$^{2+}$ ion 
[$(3d)^9$] is regarded as one $3d$ hole per Cu$^{2+}$. 
This $3d$ hole is localized on the Cu$^{2+}$ ion at $x=0$, 
and gradually acquires itinerancy with additional hole doping, 
which is understood in terms of a Kondo coupling 
with the itinerant hole on the O$^{2-}$ ion 
known as the Zhang-Rice singlet.\cite{zhang1988} 
In this sense, a hole on a Cu$^{2+}$ ion in various copper oxides 
is in the verge of the localization; 
It acts as a local moment in some materials, while it acts 
as an itinerant carrier in others.

\begin{figure}
 \includegraphics[width=8cm,keepaspectratio=true]{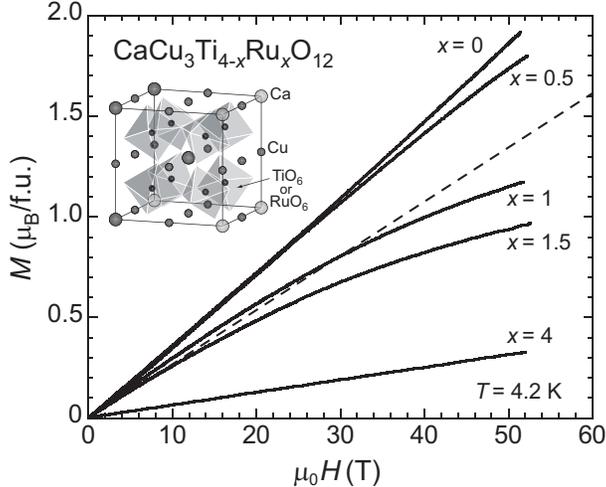} 
 \caption{
 High-field magnetization of CaCu$_{3}$Ti$_{4-x}$Ru$_{x}$O$_{12}$ 
 ($x$~=~0,~0.5~1,~1.5,~and 4) at 4.2~K. 
 The broken line is a linear fit to the data of $x=1.5$ 
 in low magnetic fields.
 In the inset the crystal structure
 of CaCu$_{3}$Ti$_{4-x}$Ru$_{x}$O$_{12}$ is schematically shown.
 }
 \label{fig01}
\end{figure}

Recently, a similar localization-itinerancy crossover 
of the Cu$^{2+}$ hole  
has been reported in the complex copper oxide 
CaCu$_{3}$Ti$_{4-x}$Ru$_{x}$O$_{12}$.\cite{kobayashi2004ccro,%
ramirez2004,tsukada2009}. 
This material crystallizes in the $AA'_3B_4$O$_{12}$ type ordered 
perovskite structure with $Im\bar{3}$ symmetry, 
in which the Ca$^{2+}$ and Cu$^{2+}$ ions 
occupy the $A$ and $A'$ sites to form 
a doubly periodic unit cell from the primitive perovskite cell 
(see the inset of Fig. \ref{fig01}). 
In this system, the ground state runs from an antiferromagnetic 
insulator at $x=0$ to a paramagnetic metal at $x=4$ 
through a spin-glass like insulator at $x=0.5-1.5$. 
This clearly indicates that 
the Cu$^{2+}$ hole systematically changes 
from a local moment in $x=0$ to an itinerant carrier in $x=4$. 
One significant difference from La$_{2-x}$Sr$_x$CuO$_4$ 
is that the formal valence of the Cu ion is always $2+$ 
in CaCu$_{3}$Ti$_{4-x}$Ru$_{x}$O$_{12}$. 
Since the substitution of Ru$^{4+}$ [$(4d)^4$] 
for Ti$^{4+}$ [$(3d)^0$] is isovalent substitution, 
the valence of the Cu ion 
is not perturbed by the Ru content $x$. 
In this sense, the crossover from localization to itinerancy 
can be more clearly investigated in the present system. 

The $x=0$ phase, i.e., CaCu$_{3}$Ti$_{4}$O$_{12}$ 
shows a huge dielectric constant of $10^4$ 
at room temperature \cite{subramanian2000} 
and an antiferromagnetic order below 25 K.\cite{kim2002} 
Although the mechanism of the large dielectric constant is not yet 
clarified, majority in the community think that it comes from an extrinsic 
origin.\cite{he2002,lunkenheimer2002,chung2004,mitsugi2007} 
Nevertheless this dielectric property is 
interesting, because this extrinsic nature can be artificially 
controlled to improve the dielectric properties.\cite{kobayashi2005ccto} 
The magnetism of this material is also interesting, and 
the ground state changes from antiferromagnetism to ferromagnetism 
by properly choosing the $B$ site ions.\cite{shiraki2007,shimakawa2008} 

The other end member of $x=4$, i.e., 
CaCu$_{3}$Ru$_{4}$O$_{12}$ is a paramagnetic metal, 
where the Cu$^{2+}$ ion loses the local moment.\cite{subramanian2002} 
This metallic compound shows substantial electron specific 
heat and Fermi-liquid-like resistivity 
at low temperatures,\cite{kobayashi2004ccro,ramirez2004} 
which roughly satisfy the Kadowaki-Woods relation. 
From the above results, Kobayashi et al.\cite{kobayashi2004ccro} 
proposed that CaCu$_{3}$Ru$_{4}$O$_{12}$ is a 
kind of heavy-fermion or valence-fluctuation system, 
which is examined in photoemission experiments.\cite{tran2006,sudayama2009} 
This picture has been further examined from 
first-principle calculation,\cite{xiang2007} 
magnetic resonance,\cite{kato2009,krimmel2009} 
and chemical substitution.\cite{maeno2009} 
A possibility of the non-Fermi-liquid behaviour 
is discussed below 1 K, and is compared with 
other heavy fermion compounds.\cite{krimmel2008,buttgen2010} 
The concept of the delocalization of the Cu$^{2+}$ hole 
has also been examined in the related compounds such as 
$L_{2/3}$Cu$_3$(Ti,Ru)$_{4}$O$_{12}$,\cite{terasaki2010b} 
CaCu$_{3}$Co$_{4}$O$_{12}$,\cite{yamada2010} 
CaCu$_{3}$Ru$_{4-x}$Mn$_{x}$O$_{12}$,\cite{calle2011} 
and CaCu$_{3}$V$_{4}$O$_{12}$.\cite{shiraki2008,morita2010}

We emphasize that response to high magnetic fields 
is particularly important in the present system.\cite{kida2010} 
In the antiferromagnetic phase of $x=0$, 
the exchange energy is evaluated to be 4 meV,\cite{kim2002} 
which is much smaller than that 
of La$_{2-x}$Sr$_x$CuO$_4$ (100 meV).\cite{takagi1989} 
It should be emphasized that 
this energy scale is comparable 
with the Zeeman energy of 40 T for an isolated $S=1/2$ spin, 
which is achievable with a non-destructive pulsed magnet. 
Thus this material offers a good playground 
to study how the external fields modify 
the interplay between localization and itinerancy 
of the Cu$^{2+}$ hole. 
Here we show the magnetic and transport properties 
of CaCu$_{3}$Ti$_{4-x}$Ru$_{x}$O$_{12}$ 
in high magnetic fields of up to 52 T. 
The measured magnetization in the intermediate phase 
($0<x<4$) is nonlinear, which can be associated with 
the field-driven delocalization of the Cu$^{2+}$ hole 
in this system. 
This concept is at least qualitatively consistent with 
a large negative magnetoresistance of $-70$\% 
observed for $x=1.5$ at 52 T at 4.2 K. 

%
\section{Experimental}
Polycrystalline samples of CaCu$_{3}$Ti$_{4-x}$Ru$_{x}$O$_{12}$ 
($x$~=~0,~0.5~1,~1.5,~and 4) 
were prepared by a solid-state reaction.\cite{kobayashi2004ccro} 
High-field magnetization and magnetoresistance measurements 
in magnetic fields of up to 52~T from 4.2 to 20~K 
were performed with non-destructive pulsed magnets at 
KYOKUGEN, Osaka University.\cite{hagiwara2006} 
The magnetization was measured by an induction method 
with a standard pick-up coil system. 
Magnetization data were obtained on a powder of the polycrystalline sample. 
The magnetoresistance was measured using a conventional 
dc four-probe technique. 
The magnetic field was applied parallel to the electrical 
current direction (longitudinal geometry).

%
\section{Results and Discussion}
High-field magnetization at 4.2~K 
is shown as a function of external field in Fig. \ref{fig01}. 
The magnetization systematically decreases 
with increasing Ru content $x$, 
which is consistent with the previous reports where 
the low-temperature susceptibility systematically changes 
from a Curie-Weiss-like to Pauli-paramagnetic-like susceptibility 
with increasing $x$.\cite{kobayashi2004ccro,ramirez2004,tsukada2009} 
When the Cu$^{2+}$ hole is fully localized, 
we expect that the saturated magnetization will 
be about 3 $\mu_{\rm B}$ per formula unit, 
corresponding to a classical $S=1/2$ spin per Cu. 
Indeed, this classical value is experimentally observed; 
Kim et al. \cite{kim2002} measured neutron diffraction of 
the antiferromagnetic order for CaCu$_3$Ti$_4$O$_{12}$, 
and found that the ordered moment is approximately 
0.8 $\mu_{\rm B}$ per Cu. 
The magnetization at 52 T is 1.9$\mu_{\rm B}$ per formula unit, 
which is two-thirds of the fully polarized moment. 
In this respect, we can understand that 
the magnetization is linear in external field of up to 52 T. 
On the other hand, the susceptibility of CaCu$_3$Ru$_4$O$_{12}$ 
($x=4$) is almost independent of temperature, 
which is regarded as the Pauli paramagnetism of itinerant electrons. 
According to the first principle calculation, 
the conduction bands consists mainly of Ru $4d$, and 
the density of states at the Fermi level is evaluated to be 
6.7 states/eV formula unit \cite{xiang2007}. 
If we estimate the Fermi energy to be a simple 
ratio of the valence electron number to the density of states, 
the Fermi energy will be around 2.3 eV 
for 16 electrons in the valence bands 
(four $4d$ electrons per Ru) per formula unit. 
Thus the magnetization is again expected to be linear in 
external field unless the Zeeman energy exceeds the Fermi energy.

In contrast to the simple magnetization for $x=0$ and 4, 
the magnetization of the intermediate phases 
($x=0.5$, 1.0, 1.5) is highly nontrivial. 
As is clearly seen in Fig. \ref{fig01}, 
their magnetization is substantially nonlinear in external 
field, and tends to saturate in high fields. 
The initial slope of the magnetization 
is roughly equal to a simple arithmetic average of 
the magnetization of the two end phases, which implies that the 
magnetization can be understood as a simple 
mixture between $x=0$ and 4. 
However, the magnetization 
bends toward $x=4$ in high fields, and in particular, 
$dM/dH$ for $x=$1.5 is close to $dM/dH$ for $x=4$ around 52 T. 
Considering that the local moment on the Cu$^{2+}$ ion 
in $x=0$ is delocalized in $x=4$, 
we suggest that the external field drives the delocalization 
of the hole on the Cu$^{2+}$ ion.

\begin{figure}
 \includegraphics[width=8cm,keepaspectratio=true]{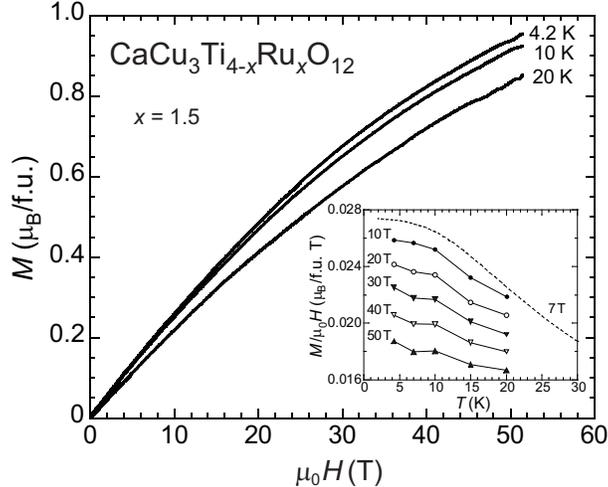} 
 \caption{
 The high-field magnetization for $x=1.5$ at various temperatures.
 The inset shows the temperature dependence of $M/\mu_0H$ 
 calculated from the isothermal magnetization. 
 A broken curve is the field-cooled susceptibility 
 at an external static field $\mu_{0}H=7$~T.
 }
 \label{fig02}
\end{figure}

Next we show the temperature dependence of 
the high-field magnetization for $x=1.5$ in Fig. \ref{fig02}. 
With increasing temperature, the magnetization gradually 
decreases, but the nonlinear behavior remains intact. 
This indicates that the field-driven delocalization works at 20 K. 
The inset of Fig. \ref{fig02} shows the temperature dependence of 
$M/\mu_{0}H$ calculated from the measured magnetization. 
A broken curve is 
the field-cooled susceptibility in a static field of 7 T 
measured with a commercial SQUID magnetometer. 
The isothermal $M/\mu_{0}H$ decreases with increasing field. 
The 7-T data look like a Curie-Weiss-like susceptibility above 10 K, 
which is consistent with the previous 
reports.\cite{kobayashi2004ccro,ramirez2004,tsukada2009} 
With increasing external fields, $M/\mu_0 H$ 
systematically decreases with weaker temperature dependence, 
and looks more Pauli-paramagnetic-like. 
This is consistent with our proposed picture that 
the external field increases a portion of the delocalized hole 
on the Cu$^{2+}$ ion. 

Before going to a phenomenological 
analysis of the nonlinear magnetization, 
we should compare the data in Figs. 1 and 2 with 
the magnetization of spin glasses that 
show similar nonlinear magnetization.\cite{majumdar1983} 
Spin glasses in alloys belong to a frustrated spin system 
in which the ferromagnetic 
and antiferromagnetic interactions compete. 
As such, the magnetic moments slowly polarize with external fields 
to show a sublinear dependence of the magnetization. 
The magnetization curves in Figs. 1 and 2 resemble 
those of spin glasses at first glance, but 
one significant difference is a typical field scale. 
Clearly, the spin glasses are easily polarized with 
a much lower magnetic field 
because ferromagnetic interaction (though partially) 
works between spins. 
On the other hand, the title compound is 
a solid solution of the antiferromagnet and paramagnet, 
and no magnetic frustration is suggested. 
Although the low-field susceptibility for $x=1.5$ 
exhibits a spin-glass-like hysteresis between zero-field-cooled 
and field-cooled processes below $T_{\rm SG}=8.5$~K,\cite{tsukada2009} 
we do not think that this hysteresis is relevant to 
the nonlinearity of the magnetization 
because it survives up to 52 T at 20 K, which is 
far larger than the energy scale of the spin-glass-like 
order of 8.5 K.

\begin{figure}
 \includegraphics[width=8cm,keepaspectratio=true]{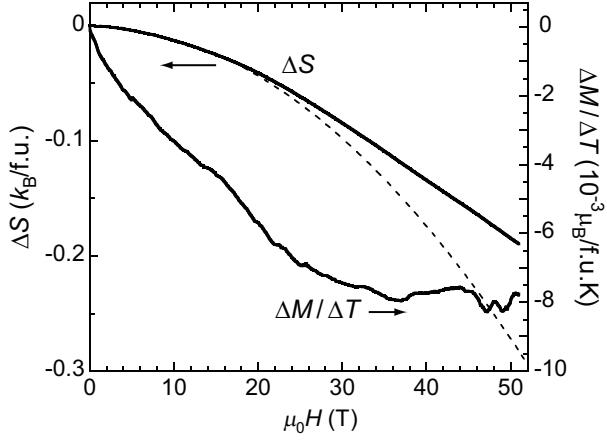}
 \caption{
 The entropy decrease ($\Delta S=\int_0^H (\partial M/\partial T) dH$) 
 plotted as a function of external field.
 The temperature dependence of the magnetization 
 $\Delta M/\Delta T=(M_{\rm 20K}-M_{\rm 10K})$/10 K 
 is also plotted as a function
 of external field, 
 which is identified to $\partial M/\partial T$.
 The dotted curve represents the extrapolation of
 $\Delta S$ from low fields (see text).
}
 \label{fig03}
\end{figure}

Let us discuss a possible mechanism of the field induced 
delocalization from a viewpoint of thermodynamics. 
We start with Maxwell's relation, 
where the temperature dependence of the magnetization 
is equal to the field dependence of the entropy, 
given as 
\begin{equation}
 \left( \frac{\partial M}{\partial T}\right)_H
  = \left( \frac{\partial S}{\partial H}\right)_T.
  \label{maxwell}
\end{equation}
We can evaluate the quantity of the left hand side of 
Eq. (\ref{maxwell}) from the high-field magnetization curve 
shown in Fig. 2. 
From the inset of Fig. 2, we find that 
the magnetization roughly linearly increases 
with decreasing temperature from 20 down to 10 K, 
which justifies the assumption 
that the slope is equal to $\partial M/\partial T$. 
Figure 3 shows the temperature slope of the magnetization 
between 10 and 20 K given by 
$(M_{\rm 20K}-M_{\rm 10K})$/10 K, 
which we identify to $\partial M/\partial T$. 
$\partial M/\partial T$ exhibits 
a complicated field dependence; 
$\partial M/\partial T$ is roughly 
linear in $H$ below around 20 T, 
while it is independent of $H$ above around 30 T. 
This is already visible in the inset of Fig. 2, 
where the temperature slope at 50 T is smaller than that at 10 T. 
A crossover field is close to the field above 
which the magnetization begins to saturate. 

Using Eq. (\ref{maxwell}), we obtain the entropy change $\Delta S$ 
by integrating $\partial M/\partial T$ as 
\begin{equation}
 \Delta S(H) = \int_{0}^{H} \left( 
 \frac{\partial M}{\partial T}\right)_{H'} dH'.
\end{equation}
The value of $\Delta S$ is plotted as a function of field 
in Fig. 3, which is always negative at all the fields measured. 
This is reasonable, because an external field suppresses 
the thermal fluctuation of spins. 
In a conventional magnet, $\Delta S$ decreases in proportion 
to $H^2$, as shown by the dotted curve in Fig. 3, 
because $M$ is linear in $H$. 
In the present system, however, the entropy decrease 
is milder than that expected from the dotted curve, 
which implies that additional entropy is induced by magnetic fields. 

This additional entropy can be explained 
in terms of the field-induced delocalization. 
When the one hole is completely localized on the Cu$^{2+}$ ion, 
the maximum entropy is $k_{\rm B} \ln 2$ from the spin degrees of freedom. 
If it is delocalized, it can move along the Cu-O-Ru network. 
Since the resistivity is high and the Cu-O-Ru network is 
highly disordered, the charge transport will be dominated by 
a variable-range-hopping process, 
where the mobile hole acquires the entropy from 
the hopping site configuration. 
A well-known case is the Heikes formula,\cite{chaikin1976} 
which gives the entropy per carrier as 
\begin{equation}
 k_{\rm B} \ln\frac{2(1-p)}{p},
\end{equation}
where $p$ is the ratio of the carrier number 
to the site number available for hopping. 
The term $\ln[(1-p)/p]$ corresponds to the configuration entropy 
which is added to the spin entropy of $k_{\rm B} \ln 2$. 
This additional entropy can be associated with 
the zero-field entropy at high temperature observed in this compound. 
Tsukada et al. \cite{tsukada2009} found that the electronic entropy for 
$x=0.5$ and 1.5 exceeds 3$k_{\rm B}\ln$2 (the maximum spin entropy of 
Cu$^{2+}$) above 60 K. 
In particular, the $x=0.5$ sample shows a significantly large entropy, 
which is unlikely to come from the doped Ru 4d electrons alone. 
Such behavior cannot be understood without the charge degrees of freedom 
of the Cu$^{2+}$ holes.

Based on the entropy argument above, 
we will propose an origin of the field-induced delocalization. 
Usually, magnetic fields favor a phase having a larger magnetization 
because the system gains the internal energy $-MdH$  with field. 
In the present case, the antiferromagnetic phase has 
a larger magnetization, and the magnetic field is expected to 
stabilize this phase. 
Thus one can expect field-induced {\it localization}, 
which is incompatible with the observation in Fig. 1. 
We suggest that the field-induced delocalization 
is basically entropy-driven. 
When the entropy gain $-SdT$ overcomes 
the internal energy loss $-MdH$, 
the system favors a more entropic phase. 
Of course, this discussion is purely 
phenomenological; 
The microscopic origin should be 
clarified by microscopic measurements 
such as magnetic resonance, photoemission spectroscopy.

\begin{figure}
 \includegraphics[width=8cm,keepaspectratio=true]{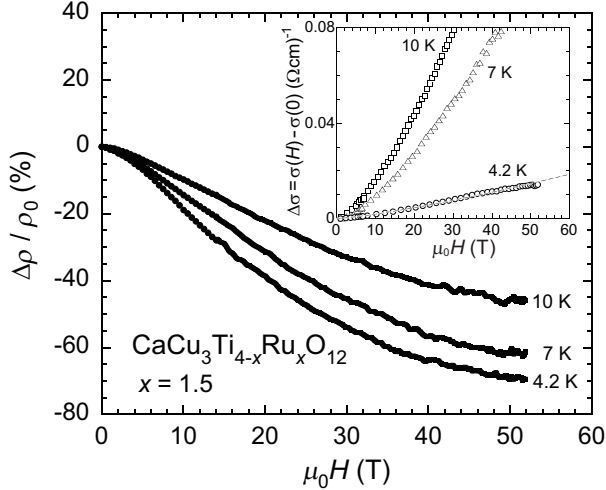} 
 \caption{
 High-field magnetoresistance of 
 CaCu$_{3}$Ti$_{4-x}$Ru$_{x}$O$_{12}$ ($x=1.5$) at 4.2,~7, and~10~K. 
 The inset shows the field dependence of 
 $\Delta \sigma = \sigma(H)-\sigma(0)$.
 }
 \label{fig04}
\end{figure}

Next let us examine the field-driven delocalization 
of the Cu$^{2+}$ hole in the charge sector. 
Figure \ref{fig04} shows the high-field magnetoresistance 
($\Delta \rho/\rho_{0}$) as a function of external field 
at 4.2,~7, and 10~K for $x=1.5$. 
A large negative magnetoresistance of 70\% is observed at 4.2 K. 
We measured the magnetoresistance  of the same sample 
in static magnetic fields of up to 14~T, and  
found that the magnetoresistance is isotropic between 
the longitudinal and the transverse configurations (not shown). 
This suggests that the large magnetoresistance does not 
arise from the orbital contribution due to the Lorentz force, 
but from the spin contribution.

Since the magnetoresistance is a second-order effect 
between conduction electrons and external fields, 
various intermediate processes are possible, 
which makes it difficult to clarify their origin experimentally. 
Therefore let us suffice it to say that 
the magnetoresistance observed here is difficult to be associated 
with any origins known thus far. 
Main causes of negative magnetoresistance are 
(i) ferromagnetic interaction/fluctuation,\cite{smit1951,majumdar1983,nigam1983} 
(ii) double exchange interaction,\cite{tokura1994,ohno1998} 
(iii) the Kondo coupling,\cite{sekitani2003} 
and (iv) localization with strong spin-orbit 
or impurity spin scattering.\cite{hikami1980,kawabata1980,rosenbaum1983} 
Clearly we can exclude the possibilities of (i) and (ii), 
because there is no trace of ferromagnetic interaction 
in the present system. 
As for (iii), we expect  $\ln H$ dependence for the 
magnetoresistance together with  $\ln T$ dependence 
for the zero-field resistance, which is not observed 
experimentally. 
The possibility of (iv) is also excluded by 
comparing the magnetoconductivity 
$\Delta\sigma$ shown in the inset of Fig. \ref{fig04} 
with theoretical calculation. 
According to the weak localization theory, 
$\Delta\sigma$ should be proportional to $\sqrt{H}$ and 
the magnitude increases with decreasing temperature,\cite{rosenbaum1983} 
which is highly incompatible with the measured data. 

Finally we will make a brief comment on 
the relationship of the negative magnetoresistance to 
the field-driven delocalization picture. 
When the magnetic fields 
delocalize the Cu$^{2+}$ holes, they 
can increase the carrier concentration $\Delta n$ 
and concomitantly increase the conductivity as 
$\Delta\sigma \sim \Delta n$. 
This is rarely seen in the magnetoresistance of 
other disordered materials, 
and the magnetic fields usually modify the mobility 
as $\Delta\sigma \sim \Delta \mu$. 
Accordingly the observed magnetoconductivity includes 
the two contributions as $n\Delta\mu + \mu\Delta n$, 
which cannot be separated at this stage. 
Thus the origin of the magnetoconductivity should be 
further investigated by measuring other transport parameters 
such as the Hall and Seebeck coefficients in high fields. 

%
\section{Summary}
In summary, we have investigated the high-field magnetization 
and magnetoresistance of the  $A$-site ordered perovskite oxide 
CaCu$_{3}$Ti$_{4-x}$Ru$_{x}$O$_{12}$~($0 \le x \le 4$). 
The magnetization for the end phases ($x=0$ and 4) is linear in external 
field, whereas that for $0.5 \le x \le 1.5$ tends to saturate in high fields. 
For $x=1.5$, a large negative magnetoresistance of $-70$\% is observed. 
We have qualitatively explained 
the nonlinear magnetization and the negative 
magnetoresistance, and have proposed a possible mechanism of 
the field-induced delocalization, 
where the holes on the Cu$^{2+}$ ions 
are delocalized in order to 
maximize the entropy related to the hopping site configuration. 
The present study has revealed that the hole on the Cu$^{2+}$ ion 
in CaCu$_{3}$Ti$_{4-x}$Ru$_{x}$O$_{12}$ is in the verge 
of localization, which is partially delocalized by external fields. 
Since the argument given here is based on the thermodynamics, 
this is independent of the detailed information of the system. 
To clarify a microscopic origin for the field-driven delocalization, 
a precise structure analysis and 
a microscopic measurements like magnetic resonance 
in high magnetic fields are indispensable. 

%
\section*{acknowledgments}
The authors would like to thank T. Mizokawa and 
Y. Shimakawa for fruitful discussion. 
This work was partly supported by 
Grant-in-Aid for Scientific Research on Priority Areas 
``New Materials Science Using Regulated Nano Spaces" (No.~19051010), 
``High Field Spin Science in 100 T" (No.~451) and 
``Invention of Anomalous Quantum Materials" (No.~16076213), 
and by a Grant-in Aid for Young Scientists (B)~(23740272) 
from the Ministry of Education, Culture, Sports, Science and Technology (MEXT), Japan.

\end{document}